\documentclass[a4paper]{article}

\newcommand{\A}{{\bf A. }}
\newcommand{\Q}{{\bf Q. }}
\newcommand{\MM}{{\cal M}} 
\newcommand{\XX}{{\cal X}} 
\newcommand{\YY}{{\cal Y}} 

\newcommand{\x}{{\em x}}
\newcommand{\y}{{\em y}}

\newcommand{\eg}{{\em eg.}}
\newcommand{\etc}{{\em etc.}}
\newcommand{\ie}{{\em i.e.}}
\newcommand{\Sch}{Schr\"{o}dinger}

\newcommand{\Ti}{I}

\begin{document}
\title{A Gravitational Explanation for Quantum Mechanics}
\author{Mark J Hadley\footnotemark \\ Dept. of Physics, Univ. of
Warwick\\  COVENTRY CV4~7AL, UK}
\date{14 September 1996}
\maketitle
\begin{abstract}
It is shown that certain structures in classical General Relativity
can give rise to non-classical logic, normally associated with Quantum
Mechanics. A 4-geon model of an elementary particle is proposed which
is asymptotically flat, particle-like and has a non-trivial causal
structure. The usual Cauchy data are no longer sufficient to determine
a unique evolution. The measurement apparatus itself can impose
non-redundant boundary conditions. Measurements of such an object
would fail to satisfy the distributive law of classical physics. This
model reconciles General Relativity and Quantum Mechanics without the
need for Quantum Gravity. The equations of Quantum Mechanics are
unmodified but it is not universal; classical particles and waves
could exist and there is no graviton.
\end{abstract}

\bibliographystyle{unsrt}
\section{Comment}\renewcommand{\thefootnote}{\fnsymbol{footnote}}
This submission reproduces the talk I gave at the 5th UK Conference on
Conceptual and Philosophical problems in Physics held in Oxford on 10th
-14th September 1996. The content follows the talk very closely but is
hopefully more coherent - what I meant to say replaces what I did say.
In a similar vein the replies to questions are what I should have said
rather than what I actually said; in both cases it is clarity rather
than the facts or the arguments which has changed (exceptions to this
rule are given as footnotes). Full references are also
included.\footnotetext[1]{email: m.j.hadley@warwick.ac.uk} 
\section{Introduction}
I am going to give a gravitational explanation of Quantum
Mechanics. By gravitation I mean Einstein's theory of General
Relativity - the unmodified classical theory. By Quantum Mechanics I
mean the Quantum Mechanics that we all know and love. As far as I am
aware nobody has given an explanation for the origin of Quantum
Mechanics before, and certainly not in terms of an established
classical theory. What is more I will do this in 20 minutes!!
\setcounter{footnote}{2}
\section{The Route from General Relativity to Quantum Mechanics}
This diagram shows the route from General Relativity to \Sch's
equation \etc\ Quantum logic has a crucial place in the path.

\setlength{\unitlength}{1mm}
\begin{figure}[h]
\begin{picture}(180,60)(0,0)
\put(0,40){\framebox(40,20)[cc]{}}
\put(20,54){\makebox(0,0)[cc]{General}}
\put(20,46){\makebox(0,0)[cc]{Relativity}}

\put(60,40){\framebox(40,20)[cc]{}}
\put(80,54){\makebox(0,0)[cc]{Quantum}}
\put(80,46){\makebox(0,0)[cc]{Logic}}

\put(0,0){\framebox(40,20)[cc]{}}
\put(20,14){\makebox(0,0)[cc]{Hilbert}}
\put(20,6){\makebox(0,0)[cc]{Space}}

\put(60,00){\framebox(40,20)[cc]{}}
\put(80,14){\makebox(0,0)[cc]{\Sch's equation}}
\put(80,9){\makebox(0,0)[cc]{Planck's constant}}
\put(80,4){\makebox(0,0)[cc]{etc.}}

\thicklines
\put(40,50){\vector(1,0){20}}
\put(100,50){\line(1,0){10}}
\put(110,50){\vector(-3,-1){120}}
\put(40,10){\vector(1,0){20}}
\put(-10,10){\vector(1,0){10}}

\end{picture}
\caption{The route from Genial Relativity to \Sch's equation via
quantum logic.}
\label{fig:route}
\end{figure}
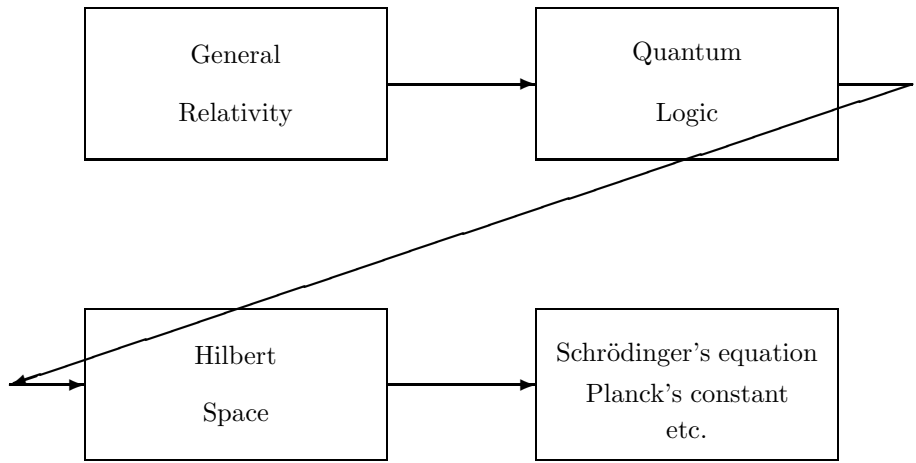

It is well known that \Sch's equation, the Dirac equation, Planck's
constant the uncertainty relations \etc, \etc\ can be derived from the
Hilbert space structure of Quantum Mechanics, the symmetries of space
and time, and the internal symmetries of the object. A good reference
to the non-relativistic case is given by Ballentine \cite{ballentine},
while Weinberg gives a useful treatment of the relativistic case
\cite{weinberg95}.

What is less well known is that the Hilbert space structure of Quantum
Mechanics is a natural representation of quantum logic. In fact it
looks increasingly as if The familiar Hilbert space structure is
unique as a vectorial representation of quantum logic\footnote{at
least for dimensions greater than 2}. Quantum logic is introduced in
the books by Jauch \cite{jauch} and Beltrametti and
Cassinelli\cite{beltrametti_cassinelli}, the latter also describes how
the Hilbert space structure is constructed from the logic.

For this talk I will show how quantum logic can arise from the
propositions (statements) about certain structures in General
Relativity. The rest is then already done for me.

\section{Quantum Logic}
Quantum logic is a non-distributive or non-Boolean logic, which means
the failure of the familiar distributive law:
\begin{equation}
a \wedge (b \vee c) \neq (a \wedge b) \vee (a \wedge
c) \label{eq:nondist}
\end{equation}
where $\wedge$ is the AND operation and $\vee$ is the OR
operation. $a$, $b$ and $c$ are the propositions or statements about
the system or state. For this talk I will use a special case of
equation~\ref{eq:nondist} - taking $c$ to be NOT $b$, denoted $ \neg
b$, and introducing the trivial operator \Ti, which is TRUE for any
state. We then have:
\begin{eqnarray}
a = a \wedge \Ti = a \wedge (b \vee \neg b) & \neq& (a \wedge b) \vee
(a \wedge \neg b) \\ 
\Rightarrow \hspace{15mm}a & \neq & (a \wedge b)
\vee (a \wedge \neg b) \label{eq:distsimple}
\end{eqnarray}

In fact quantum logic requires the distributive law to be replaced by
a weaker orthomodular condition and for a complete orthomodular
orthocomplemented atomic lattice with the covering property needs to
be constructed. This can be done. It is the subject of a paper
submitted to Foundations of Physics and of my PhD thesis. For this
talk I will only show the failure of the distributive law in the form
of equation~\ref{eq:distsimple}, because this marks the departure from
a classical system and is by itself a remarkable achievement.

\section{General Relativity}
For this work the significant features of General Relativity are:
\begin{itemize}
\item The equation, ${\bf G} = 8 \pi {\bf T}$, which relates the curvature
of spacetime, of which ${\bf G}$ is a measure, to the energy momentum
tensor ${\bf T}$.
\item It is a non-linear equation for the metric, containing first
and second derivatives and both linear and quadratic terms in the
metric.
\item The equations describe  distorted, curved spacetime.
\item The equations are local, they do not prescribe the topology,
although they may set constraints on the topology.
\item The theory allows closed timelike curves, CTCs, (just a
respectable way to say time travel). This is one of the great mysteries
of General Relativity - if CTCs are possible then how can we make them
and use them, and if not, then what forbids them. The mathematical
structure of General Relativity allows CTCs and exact solutions are
known with CTCs.
\end{itemize}

\section{CTCs}
CTCs are crucial for the results which follow, because when
interactions are allowed in spacetimes with CTCs the normal boundary
conditions are no longer adequate to uniquely determine the evolution.

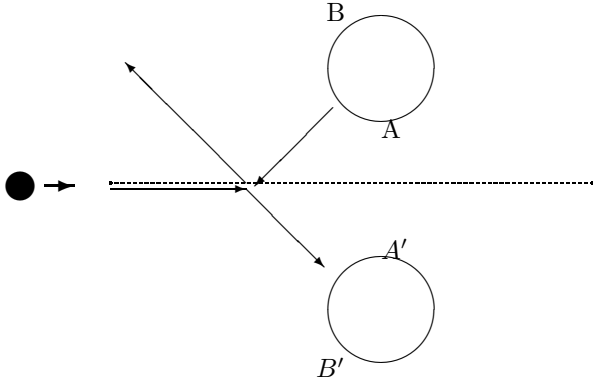
\begin{figure}[h]
\setlength{\unitlength}{.4mm}
\center{
\begin{picture}(200,120)(-100,-60)
\put(0,-40){\circle{80}}
\put(0, 40){\circle{80}}
\put(0,20){\makebox(0,0)[l]{A}}
\put(0,-20){\makebox(0,0)[l]{$A^\prime$}}
\put(-12,56){\makebox(0,0)[br]{B}}
\put(-12,-56){\makebox(0,0)[tr]{$B^\prime$}}
\put(-90,0){\vector(1,0){45}}
\put(-45,0){\vector(1,-1){26}}
\put(-16,27){\vector(-1,-1){26}}
\put(-45,2){\vector(-1,1){40}}

\put(-90,2){\dashbox{1}(160,0){}}
\put(-120,1){\circle*{10}}
\thicklines
\put(-112,1){\vector(1,0){10}}

\end{picture}}
\caption{The ball travelling from the left may be hit by itself
into one mouth of the wormhole, to emerge at an earlier time to cause
the impact.}
\label{fig:ctcholes}

\end{figure}

Consider a billiard ball in a plane, given an initial position and
velocity then the subsequent trajectory is determined, see the dashed
line in figure~\ref{fig:ctcholes}; even if there are walls, or hills,
or in this example a wormhole.
 
If the wormhole is replaced with a time-machine, so that a
particle which enters one mouth exits at a corresponding point from
the other mouth, but at an earlier time. The original trajectory
is still a possible consistent solution, but now alternatives
exist. For example the ball could be hit into the mouth of the
wormhole, reappear from the other mouth at an earlier time in such a
direction that it causes the original collision (see the solid
lines in figure~\ref{fig:ctcholes}). It must be stressed
that these are both consistent evolutions of the system even though
the initial data would normally (in the absence of CTCs) give a unique
trajectory.

The multiplicity of possible solutions is not confined to this
example. It is considered to be a generic feature when
self-interacting objects or fields are in a spacetime with CTCs (see
for example papers by Friedman {\em et al} \cite{friedman_morris} and
Thorne \cite{thorne}).

\section{4-Geon}
The strange features of CTCs are exploited in a model of an elementary
particle which I call a 4-geon. The idea that an elementary particle
is a solution of the field equations (of General Relativity or any
unified field theory) dates from the earliest days of General
Relativity. Einstein attempted to find such solutions in all his
theories. In the 60's Misner and Wheeler\cite{misner_wheeler}
continued with the work and used the term {\em geon} to describe a
topologically non-trivial spacetime structure held together by its own
gravitational attraction. However most of the earlier work used a
topologically non-trivial three-manifold evolving with time, and
assumed that a global time coordinate existed. By contrast a 4-geon
has a non-trivial causal structure. A 4-geon is assumed to have the
following properties:
\begin{itemize}
\item It is a solution of the field equations of General Relativity.
\item It has a non-trivial causal structure.
\item Interactions are taking place around CTCs.
\item The metric is asymptotically flat.
\item Particle-like: the region of non-trivial topology will be
found in one and only one place - otherwise it would not be
recognisable as a particle.
\end{itemize}

With this model of an elementary particle the normal boundary
conditions can no longer be expected to be adequate to determine a
unique evolution.
\newpage
\section{Boundary Conditions and Measurements}
The idea that the state preparation sets boundary conditions is
obvious. In our real or imagined experiments we look for outcomes
consistent with the preparation conditions; this may comprise a source,
collimators, shutters filters \etc:

\begin{figure}[h]
\setlength{\unitlength}{.01mm}
\begin{picture}(12876,5452)(-599,-5191) 

\thicklines
\put(600,-2161){\circle*{336}}
\put(1800,-361){\line( 0,-1){1575}}
\put(3000,-361){\line( 0,-1){1575}}
\put(1800,-2380){\line( 0,-1){1575}}
\put(3000,-2380){\line( 0,-1){1650}}
\put(4200,-2761){\framebox(2100,1200){Filter}}

\thinlines
\put(5251,-2911){\vector( 1,-1){1050}}
\put(826,-1936){\vector( 3, 1){675}}
\put(826,-2311){\vector( 2,-1){660}}
\put(676,-2386){\vector( 1,-1){825}}
\put(676,-1936){\vector( 1, 1){825}}
\put(600,-1861){\vector( 0, 1){900}}
\put(600,-2461){\vector( 0,-1){900}}
\put(900,-2161){\vector( 1, 0){3000}}
\put(6376,-2161){\vector( 1, 0){3000}}
\put(300,-3736){\makebox(0,0)[lb]{Source}}
\put(2400,-4411){\makebox(0,0)[cb]{Collimator}} 
\end{picture}
\caption{The boundary conditions imposed by state-preparation}
\label{fig:init}
\end{figure}
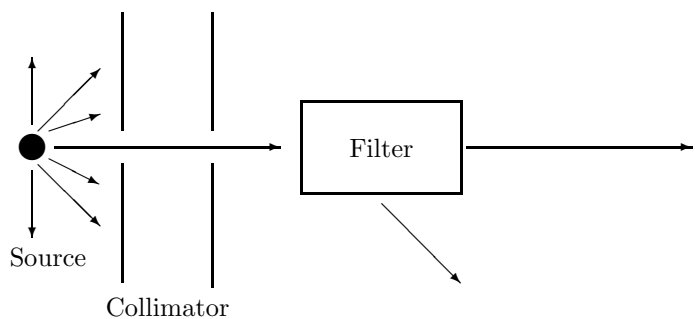

With the 4-geon model of a particle the state preparation conditions
will no longer be adequate to uniquely determine the subsequent
evolution. The measurement apparatus itself can set further boundary
conditions which are not redundant. An \x-spin measurement is an
example:

\begin{figure}[h]
\setlength{\unitlength}{.01mm}
\begin{picture}(12876,5452)(-599,-5191) 

\thicklines
\put(12000,239){\line( 0,-1){4800}}
\put(11700,239){\line( 1, 0){300}}
\put(11700,-61){\line( 1, 0){300}}
\put(11700,-361){\line( 1, 0){300}}
\put(11700,-661){\line( 1, 0){300}}
\put(11700,-961){\line( 1, 0){300}}
\put(11700,-1261){\line( 1, 0){300}}
\put(11700,-1561){\line( 1, 0){300}}
\put(11700,-1861){\line( 1, 0){300}}
\put(11700,-2161){\line( 1, 0){300}}
\put(11700,-2461){\line( 1, 0){300}}
\put(11700,-2761){\line( 1, 0){300}}
\put(11700,-3361){\line( 1, 0){300}}
\put(11700,-3661){\line( 1, 0){300}}
\put(11700,-3961){\line( 1, 0){300}}
\put(11700,-4261){\line( 1, 0){300}}
\put(11700,-4561){\line( 1, 0){300}}
\put(11700,-3061){\line( 1, 0){300}}
\put(11700,-4861){\makebox(0,0)[b]{$x$-position}}
\put(11700,-5161){\makebox(0,0)[b]{measurement}}
\put(7200,-2761){\framebox(2100,1200){}}
\put(8250,-1861){\makebox(0,0)[b]{\small $x$-oriented}}
\put(8250,-2161){\makebox(0,0)[b]{\small Stern-}}
\put(8250,-2461){\makebox(0,0)[b]{\small Gerlach}}
\put(600,-2161){\circle*{336}}
\put(1800,-361){\line( 0,-1){1575}}
\put(3000,-361){\line( 0,-1){1575}}
\put(1800,-2380){\line( 0,-1){1575}}
\put(3000,-2380){\line( 0,-1){1650}}
\put(4200,-2761){\framebox(2100,1200){Filter}}

\put(8100,-4861){\vector( 0, 1){600}}
\put(8100,-4861){\vector( 1, 0){600}}

\thinlines
\put(5251,-2911){\vector( 1,-1){1050}}
\put(826,-1936){\vector( 3, 1){675}}
\put(826,-2311){\vector( 2,-1){660}}
\put(676,-2386){\vector( 1,-1){825}}
\put(676,-1936){\vector( 1, 1){825}}
\put(600,-1861){\vector( 0, 1){900}}
\put(600,-2461){\vector( 0,-1){900}}
\put(900,-2161){\vector( 1, 0){3000}}
\put(9600,-2461){\vector( 3,-2){1453.846}}
\put(9600,-1861){\vector( 3, 2){1453.846}}
\put(6376,-2161){\vector( 1, 0){675}}
\put(300,-3736){\makebox(0,0)[lb]{Source}}
\put(2400,-4411){\makebox(0,0)[cb]{Collimator}} 
\put(8400,-5161){\makebox(0,0)[b]{$z$}}
\put(7800,-4561){\makebox(0,0)[b]{$x$}}
\end{picture}
\caption{The boundary conditions imposed by state-preparation and an
$x$-spin measurement}
\label{fig:xmeas}
\end{figure}
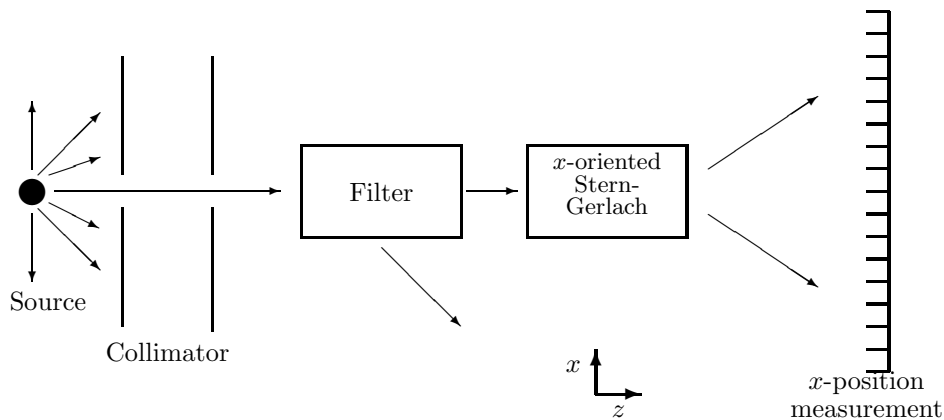

Note that the measurement apparatus is physically very similar to the
state preparation. Within the structure of the 4-geon there can be a
causal link between the measurement apparatus, state preparation and
the evolution, which gives a physical explanation for
measurement-dependent effects.

Alternatively we could measure the \y-spin with a very similar
apparatus. However, as is well known, a \y-oriented Stern-Gerlach
filter and an \x-oriented one are physically incompatible. They set
conflicting boundary conditions.

\begin{figure}[h]
\setlength{\unitlength}{.01mm}
\begin{picture}(12876,5452)(-599,-5191) 

\thicklines
\put(7200,-2761){\framebox(2100,1200){}}
\put(600,-2161){\circle*{336}}
\put(1800,-361){\line( 0,-1){1575}}
\put(3000,-361){\line( 0,-1){1575}}
\put(1800,-2380){\line( 0,-1){1575}}
\put(3000,-2380){\line( 0,-1){1650}}
\put(4200,-2761){\framebox(2100,1200){Filter}}

\put(11700,-3961){\line( 1, 0){600}}
\put(10200,-1561){\line( 1, 0){375}}
\put(10426,-1861){\line( 1, 0){375}}
\put(9889,-656){\line( 3,-4){2448}}
\put(11551,-3661){\line( 1, 0){525}}
\put(11400,-3361){\line( 1, 0){450}}
\put(11251,-3061){\line( 1, 0){375}}
\put(11026,-2761){\line( 1, 0){375}}
\put(10876,-2461){\line( 1, 0){300}}
\put(10651,-2161){\line( 1, 0){300}}
\put(9600,-661){\line( 1, 0){300}}
\put(9826,-961){\line( 1, 0){300}}
\put(9976,-1261){\line( 1, 0){375}}

\put(11026,-2161){\vector( 0, 1){600}}
\put(11026,-2161){\vector( 1, 0){600}}
\put(11026,-2176){\vector( 3,-4){340}}

\thinlines
\put(5251,-2911){\vector( 1,-1){1050}}
\put(826,-1936){\vector( 3, 1){675}}
\put(826,-2311){\vector( 2,-1){660}}
\put(676,-2386){\vector( 1,-1){825}}
\put(676,-1936){\vector( 1, 1){825}}
\put(600,-1861){\vector( 0, 1){900}}
\put(600,-2461){\vector( 0,-1){900}}
\put(900,-2161){\vector( 1, 0){3000}}
\put(6376,-2161){\vector( 1, 0){675}}

\put(9500,-1951){\vector( 1, 1){400}}
\put(9500,-2311){\vector( 3,-2){1400}}

\put(300,-3736){\makebox(0,0)[lb]{Source}}
\put(2400,-4411){\makebox(0,0)[cb]{Collimator}} 

\put(7200,-2761){\framebox(2100,1200){}}
\put(8250,-1861){\makebox(0,0)[b]{\small $y$-oriented}}
\put(8250,-2161){\makebox(0,0)[b]{\small Stern-}}
\put(8250,-2461){\makebox(0,0)[b]{\small Gerlach}}

\put(11851,-2161){\makebox(0,0)[b]{$z$}}
\put(11251,-1636){\makebox(0,0)[b]{$x$}}
\put(11476,-2566){\makebox(0,0)[lb]{$y$}}
\put(11896,-4396){\makebox(0,0)[b]{$y$-position}}
\put(11911,-4681){\makebox(0,0)[b]{measurement}}

\end{picture}
\caption{The boundary conditions imposed by state-preparation and a
$y$-spin measurement}
\label{fig:ymeas}
\end{figure}
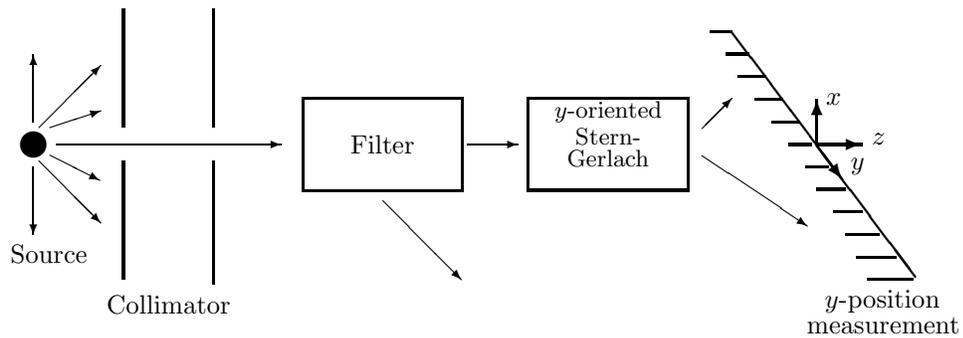

\newpage
\section{Sets of Manifolds and Propositions}
To see the effect of these boundary conditions we will consider the
possible sets of 4-geon manifolds.  Let $\MM$ denote the set of
4-manifolds consistent with the state preparation conditions. While
$\XX$ denotes those manifolds consistent with both the
state-preparation and an \x-spin measurement. $\XX$ is partitioned
into the two disjoint subsets $\XX^+$ and $\XX^-$.

\setlength{\unitlength}{0.8mm}
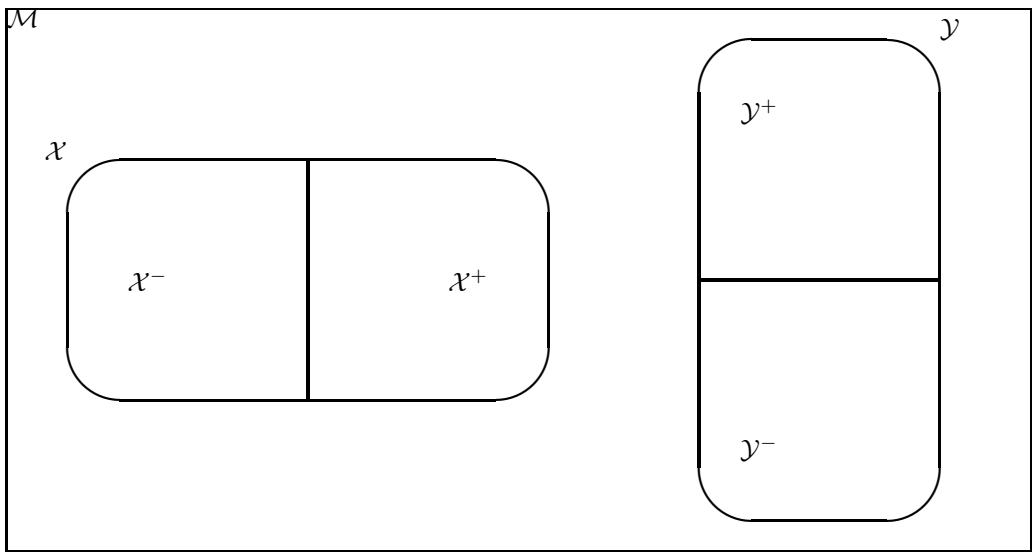
\begin{figure}[h]
\begin{picture}(220,100)(15,0)

\put(5,5){\framebox(170,90)[tl]{$\MM$}}

\thicklines

\put(55,50){\oval(80,40)}
\put(15,70){\makebox(0,0)[br]{$\XX$}}
\put(55,30){\line(0,1){40} }
\put(25,50){\makebox(0,0)[l]{$\XX^-$}}
\put(85,50){\makebox(0,0)[r]{$\XX^+$}}

\put(140,50){\oval(40,80)}
\put(160,90){\makebox(0,0)[bl]{$\YY$}}
\put(120,50){\line(1,0){40} }
\put(130,20){\makebox(0,0)[b]{${\bf \YY^-} $}}
\put(130,80){\makebox(0,0)[t]{${\bf \YY^+} $}}

\thinlines
\end{picture}
\caption{Sets of 4-manifolds consistent with both state preparation
and the boundary conditions imposed by different measurement
conditions.}
\label{fig:venn}
\end{figure}

This simple diagram is immediately non-classical, because classically
the measurement must simply partition those solutions $\MM$ consistent
with the state preparation; it cannot define a proper subset of $\MM$.

The \y-measurement defines a different subset of $\MM$, denoted $\YY$,
which is disjoint from $\XX$. 

The propositions are statements about the state preparation; they are
not in one to one correspondence with the measurements because some
measurements give the same information about the state (they are
indistinguishable by any state preparation). For this system, $\XX$
corresponds to {\em there is a manifold in $\MM$ consistent with an
\x-spin measurement} and $\YY$ to {\em there is a manifold in $\MM$
consistent with a \y-spin measurement} these are both the trivial
proposition which is always true and which we denote by \Ti.

By contrast $\XX^+ \cap \YY^+$ corresponds to {\em there is a manifold
in $\MM$ consistent with a positive \x-spin measurement and also with
a positive \y-spin measurement}, but $\XX^+$ and $\YY^+$ are disjoint
and so the intersection corresponds to the trivial proposition which
is always false $\XX^+ \cap \YY^+ = \emptyset$. So we have:
\begin{equation}
\XX \neq  (\XX^+ \cap \YY^+) \cup (\XX^+ \cap \YY^-)
\end{equation}
which is the failure of the distributive law for the propositions.

\section{Summary}
The conjectured 4-geon description of particles is speculative. I
cannot produce a solution of the field equations with the required
properties. I have not tried to. The advice I have received is not to
try and find a solution because it is so difficult so solve Einstein's
equations, especially if solutions are highly non-linear, lacking in
symmetry and topologically non-trivial.

However, In other respects this theory is extremely conservative - it keeps
General Relativity in its unmodified form and it retains 3+1 dimensions
for space and time.

The unifying nature of the theory justifies the speculation. Field and
particle descriptions of Nature are unified as Einstein had always
hoped and expected. For the first time the origin of Quantum Mechanics
is explained in terms of existing theories. In doing so, General
Relativity and Quantum Mechanics are reconciled, not with a quantum
theory of gravitation as was expected, but with a gravitational
explanation for Quantum Mechanics. There is no simpler or more
conservative theory which reconciles Quantum Mechanics and General
Relativity.
 
\section{Predictions}
Despite giving standard Quantum Mechanics with the same equations and
structure the theory does make some new predictions:
\begin{itemize}
\item There is no quantum theory of gravity.
\item Classical objects are possible. The peculiar 4-geon structures
give rise to quantum effects; if these are absent then classical
deterministic evolution would occur.
\item There is no graviton. This follows from either of the statements
above. Gravitational waves are topologically simple solutions of
Einstein's equations without CTCs. Therefore they cannot exhibit
quantum phenomena such as wave particle duality. Gravitational waves
are not quantised.
\end{itemize}

\section{Questions}
The following questions were asked in open discussions or
afterwards. They were most helpful to me and I thank all those who
joined in. Apologies for not giving names and for any errors, but I
did not make notes at the time.

\Q You have shown the failure of the distributive law, but for quantum
logic you must show much more - orthomodularity, atomicity \etc\ Can
you show this too?
 
\A Yes, the failure of the distributive law is the most remarkable
feature because it marks the divergence of classical and non-classical
systems. Orthomodularity can be shown\cite{mjh_found}, in fact it
follows easily since this construction relies upon the measurement
apparatus and so the arguments of Mackey (see
\cite{beltrametti_cassinelli}[page~147]) apply. Atomicity is a
mathematical idealisation which cannot be derived, but this sort of
idealisation is common to all of mathematical physics; \eg\ the use of
real numbers to represent momentum in classical physics. 

\Q How can spin-half arise in a gravitational theory?

\A There is an enormous richness in the choice of topology. Certain
manifolds can be shown to have the transformation properties of a
spinor provided a fixed asymptotically flat background metric is
assumed. See the fascinating paper by Friedman and Sorkin
\cite{friedman_sorkin} or the discussion in my thesis).

\Q How does the superposition of the wavefunction arise in this model?

\A The wavefunction just gives information about the probability
measurement outcomes. If the logic were Boolean a real number between
0 and 1 would suffice and there would be only trivial
superpositions. However to represent probabilities for a non-Boolean
logic, complex-valued wavefunctions are required.

\Q How can you get Quantum Mechanics which is formulated on a flat
spacetime when you are considering manifolds with a nontrivial
topology?

\A The manifolds are asymptotically flat. Quantum Mechanics can be
regarded as a way of mapping information about these knots of
spacetime onto the flat spacetime which we are familiar with. It is in
the asymptotically flat region that we set boundary conditions \etc\

\Q Do solutions of Einstein's equations exist with CTCs which can be
traversed in a finite time?

\A Yes. 

\Q There are alternative ways of assigning probabilities to an
orthomodular lattice which cannot be represented by a Hilbert
space. Consider for example the model by Mielnik\cite{mielnik} which
can be found in \cite{beltrametti_cassinelli}[page~205].

\A I am not aware of that example. However the very existence of
non-classical logic in systems described by a classical theory is by
itself most remarkable, to get quantum logic as well is amazing. The
familiar Hilbert space structure is certainly compatible with this logic even
if it is not unique.

\Q Are you aware of other work in which an orthomodular lattice is
constructed geometrically from subsets of flat Minkowski space?

\A I have seen geometric 
constructions of orthomodular lattices \eg\
Watanabe \cite{watanabe}[page~303]. I think that
such models rely on an innovative definition of complementation, they
are interesting but not particularly remarkable. My
work shows that the orthocomplemented lattice arises with the
definitions of complementation associated with real experiments. In
this respect the construction is unique.

\Q What is the energy tensor responsible for the spacetime knots?

\A I have deliberately not made assumptions about the energy -momentum
tensor. The most appealing case would be for it be zero \ie\ a vacuum
solution. Spacetime can be knotted without any
source. Indeed the wormholes (not traversable ones) can be 
solutions of the source-free field equations.

\section{Acknowledgements}
I would like to thank Harvey Brown, Mauricio Suarez and Katherine
Brading from the Department of Philosophy, Oxford University, for
organising the conference. This work has been supported by the
University of Warwick.

\end{document}